\documentclass[twocolumn,amsmath,ammsymb]{revtex4-1}
\usepackage{ulem}
\usepackage{epsfig}
\usepackage{amsmath}
\usepackage{amssymb}
\usepackage{graphicx}
\usepackage{dcolumn}
\usepackage{soul}
\usepackage{bm}
\usepackage[usenames]{color}
\usepackage[breaklinks]{hyperref}
\hypersetup{colorlinks=true,allcolors=blue}

\begin{document}

\preprint{}

\title{Examining the validity of the two-dimensional conical model to describe the three-dimensional ZrTe$_5$}
\author{Yi-Xiang Wang$^1$ and Fuxiang Li$^2$}
\affiliation{$^1$School of Science, Jiangnan University, Wuxi 214122, China.}
\affiliation{$^2$School of Physics and Electronics, Hunan University, Changsha 410082, China}

\date{\today}

\begin{abstract}
Understanding the low-energy excitation state in three-dimensional (3D) layered compound ZrTe$_5$ remains a challenging problem in the study of novel topological materials.  Recently a two-dimensional conical model was proposed to explain the experimental optical spectroscopy in 3D ZrTe$_5$ [Phys. Rev. Lett. 122, 217402 (2019)].  Motivated by this work, in this paper, we perform a systematic theoretical study on the optical conductivity of this model in both cases without and with an external magnetic field, in order to further demonstrate the validity of this model and to recover new physics.  We find that there exist completely different characteristics for optical conductivity along different directions, due to anisotropic low-energy excitations in this two-dimensional conical model.  Specifically, for the interband optical conductivity, we find asymptotic dependence on the optical frequency as Re$(\sigma_x)\sim\omega^{\frac{1}{2}}$ and Re$(\sigma_z)\sim\omega^{\frac{3}{2}}$, which are universal both in the gapped insulator phase and Weyl semimetal phase.  For the magneto-optical conductivity, on the contrary,  Re$(\sigma_{x/z}^B)$ shows distinct signatures in the gapped insulator phase and Weyl semimetal phase, which can help distinguish the two phases.  Our results, to be verified in future experiments, could provide more insights in the understanding of topological nature in ZrTe$_5$.
\end{abstract}

\maketitle

\section{Introduction}

Dirac semimetal stands as a paradigmatic representative of a symmetry-protected gapless topological state.  It can be realized in the low-energy excitation of the pristine two-dimensional (2D) graphene~\cite{A.H.C.Neto} and also in several three-dimensional (3D) materials~\cite{N.P.Armitage}, such as Cd$_3$As$_2$~\cite{S.Borisenko, Z.K.Liua, M.Neupane} and Na$_3$Bi~\cite{Z.K.Liub}.  The 3D Dirac node can be regarded as two overlapped Weyl nodes of opposite chiralities and is protected by the time-reversal and the spatial inversion symmetry.  When either symmetry is broken, the Dirac node splits into a pair of Weyl nodes and the system comes into the so-called Weyl semimetal (WSM) phase, which has been found to exist in a multilayer heterostructure~\cite{A.A.Burkov} and TaAs family~\cite{B.Q.Lv, L.X.Yang, S.Y.Xu}. 

Among the studies of 3D topological materials, a layered compound ZrTe$_5$, with an extremely high mobility, has aroused many interests~\cite{H.Weng, R.Y.Chena, R.Y.Chenb, Z.G.Chen, L.Moreschini, G.Manzoni, B.Xu2018, Y.Liu, Q.Li, X.B.Li, R.Wu, E.Martino}.  Both theoretical and experimental results support a nontrivial topology of the low-energy excitation states in ZrTe$_5$.  It is, however, still under heated debate so far,  regarding the nature of the topological character.  The early \textit{ab initio} calculations indicated that ZrTe$_5$ was close to the phase boundary between the strong and weak topological insulators (TIs)~\cite{H.Weng}, of which only the former has topologically protected surface states~\cite{M.Z.Hasan}.  The angle-resolved photoemission spectroscopy (ARPES)~\cite{Q.Li}, infrared spectroscopy~\cite{R.Y.Chena}, magneto-optical~\cite{R.Y.Chenb} measurements were interpreted in terms of the 3D gapless Dirac semimetal.  The scanning tunneling microscopy (STM) studies~\cite{X.B.Li, R.Wu} found that ZrTe$_5$ was a weak 3D TI with a bulk band gap of about 80-100 meV, which was also supported by an ARPES study of the evolution of the band structure with temperature and surface doping~\cite{L.Moreschini}.  To the contrary, other ARPES and optical transition studies revealed a metallic character of the sample surface and suggested that ZrTe$_5$ was a strong TI~\cite{G.Manzoni, Z.G.Chen, B.Xu2018}.  These contradictory verifications about the low-energy excitation state in ZrTe$_5$ may be attributed to its strong sensitivity on the details of the lattice parameter and also on the purity of the crystals. 

Recently, based on an optical spectroscopy study in ZrTe$_5$, Martino et al.~\cite{E.Martino} proposed a novel viewpoint that its low-energy excitation should be more suitably described by  a 2D conical model, as in Eq.~(\ref{model}), rather than by the 3D massless/massive Dirac model.  In this 2D conical model, the linear conical dispersion is not 3D, but only 2D in the $x-y$ plane.  A natural question arises that, is the proposed 2D conical model valid or sufficient to describe the low-energy excitation in 3D ZrTe$_5$?  To explore this model and to reveal the topological nature in ZrTe$_5$, in this paper, we perform a systematic theoretical study on the optical conductivity of the 2D conical model.  We will consider both cases without and with an external magnetic field.

Our main results are as follows.  Starting from the Kubo's formula, we obtain the analytical expressions for the optical conductivity in the clean limit and consider the effects of  impurity scatterings  by phenomenologically introducing a finite scattering rate.  We find that, due to the anisotropic low-energy states, there exist completely different characteristics of the optical conductivity along the $x-$ and $z-$direction.  (i) For the interband optical conductivity, we find that no matter what phase the system lies in, the asymptotic dependence on the photon frequency is given as Re$(\sigma_x)\sim\omega^{\frac{1}{2}}$ and Re$(\sigma_z)\sim\omega^{\frac{3}{2}}$, which are less affected by the impurity scatterings.  For the optical conductivity Re$(\sigma_x)$ in the band insulator phase, we obtain an accurate and exact expression compared to that in Ref.~\cite{E.Martino}, as the whole wave vectors in the Brillouin zone are included in our calculations of the current density operator.  (ii) For the magneto-optical conductivity, Re$(\sigma_{x/z}^B)$ exhibits distinct signatures in the gapped insulator phase and WSM phase, which can provide clear evidences to distinguish the two phases.  With increasing impurity scattering, numerics show that they can smoothen the signatures, such as the resonant peaks and kinks.  Our work may provide some guidances for experiment in the future and can help verify the low-energy excitation state in ZrTe$_5$.

\section{Energy bands and Landau Levels}

\begin{figure}
	\includegraphics[width=8.4cm]{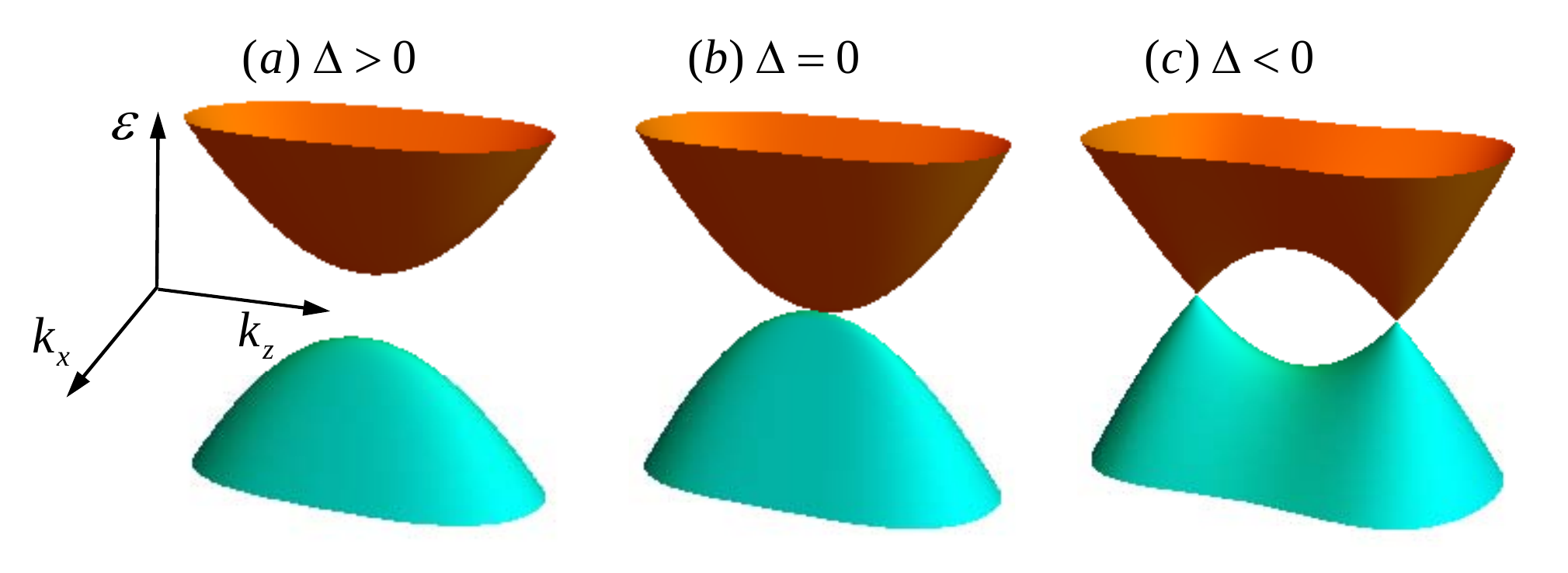}
	\caption{(Color online) Schematic plot of the energy bands described by $H(\boldsymbol k)$ in the $(k_x, k_z)$ space with $k_y=0$.  (a) the band insulator $\Delta>0$, (b) the critical phase $\Delta=0$, and (c) the WSM $\Delta<0$. }
	\label{Fig1}
\end{figure}  

We start from the description of 3D system with the 2D conical dispersion model.  The two-band Hamiltonian is given as ($\hbar=1$)~\cite{E.Martino, H.Z.Lu, D.K.Mjkherjee},  
\begin{align}
H(\boldsymbol k)=v (k_x\tau_x +k_y\tau_y)
+(\Delta+\zeta k_z^2)\tau_z,
\label{model}
\end{align}
where the Pauli matrices $\tau$ act on the pseudospin degree of freedom, such as orbital or sublattice, $v$ is the isotropic Fermi velocity in the $x-y$ plane, $\Delta$ has the dimension of energy and $\zeta=\frac{1}{2m^*}$, with $m^*$ being the effective mass in $z-$direction.  Throughout this paper, we neglect the spin-orbit coupling and assume that spin is a good quantum number.  The Hamiltonian preserves the inversion symmetry ${\mathcal I}^{-1}H(-\boldsymbol k)
{\mathcal I}=H(\boldsymbol k)$, with ${\mathcal I}=\tau_z$.  The nonvanishing $\tau_z$ term in $H(\boldsymbol k)$ breaks the time-reversal symmetry ${\mathcal T}^{-1}H(-\boldsymbol k){\mathcal T}\neq H(\boldsymbol k)$, with ${\mathcal T}=i\tau_y{\mathcal K}$ and $\mathcal K$ being the complex conjugation operator.  Thus, the system belongs to the unitary class A in the Altland and Zirnbauer notations~\cite{A.P.Schnyder, C.K.Chiu}. 

Without external magnetic field, the energy and the corresponding eigenvector of $H(\boldsymbol k)$ are obtained directly as
\begin{align}
\varepsilon_s=s\varepsilon=s\sqrt{v^2(k_x^2+k_y^2)+(\Delta+\zeta k_z^2)^2}, 
\end{align}
and 
\begin{align}
\psi_+=\begin{pmatrix} \chi_+ e^{-i\theta}\\ \chi_-
\end{pmatrix},
\psi_-=\begin{pmatrix} \chi_- e^{-i\theta}\\ -\chi_+
\end{pmatrix}, 
\end{align} 
with the band index $s=\pm1$, $\text{tan}\theta=\frac{k_y}{k_x}$, and $\chi_\pm=\sqrt{\frac{1}{2}\pm\frac{\Delta+\zeta k_z^2}{2\varepsilon}}$.  The Hamiltonian represents (i) the band insulator for $\Delta>0$, (ii) the critical phase for $\Delta=0$ and (iii) the WSM for $\Delta<0$.  In the WSM phase, the Weyl points are located at ${\boldsymbol K}_\pm=(0,0,\pm\sqrt{-\frac{\Delta}{\zeta}})$, with the sign $\pm$ indicating opposite chiralities of the two Weyl nodes.  The different phases controlled by $\Delta$ are schematically plotted in Fig.~\ref{Fig1}.  In Ref.~\cite{E.Martino}, the experimental studies supported the ZrTe$_5$ sample lies in the gapped band insulator phase, with $\Delta=3$meV derived from the 
magneto-optical transmission measurements.  We note that a two-dimensional analogue of $H(\boldsymbol k)$ was used to study the quantum multicriticality near the critical point driven by the short-range interactions~\cite{B.Roy}.

When a magnetic field is present in the system, we assume it along the $z-$direction, ${\boldsymbol B}=(0,0,B)$.  We choose the Landau gauge ${\boldsymbol A}=(-yB,0,0)$ and make the Periels substitution $\boldsymbol p\rightarrow\boldsymbol p-e\boldsymbol A$.  By using the raising and lowering operators, the energy and eigenvector of the $n-$LL are obtained as: 
\begin{align}
&\varepsilon_{n>1,s}=s\varepsilon_n=s\sqrt{2n v^2 l_B^{-2}+(\Delta+\zeta k_z^2)^2}, 
\\
&\varepsilon_0=\Delta+\zeta k_z^2, 
\end{align}
and 
\begin{align}
\psi_{n+}=\begin{pmatrix}
\chi_{n+}\phi_n\\ \chi_{n-}\phi_{n-1} 
\end{pmatrix},  
\psi_{n-}=\begin{pmatrix}
\chi_{n-}\phi_n\\-\chi_{n+}\phi_{n-1}
\end{pmatrix},  
\psi_0=\begin{pmatrix}
\phi_0\\ 0
\end{pmatrix}, 
\end{align}
where $l_B=\sqrt{\frac{1}{eB}}=\frac{25.6\text{nm}}{\sqrt B}$ is the magnetic length, $\chi_{n\pm}=\sqrt{\frac{1}{2}\pm\frac{\Delta+\zeta k_z^2}{2\varepsilon_n}}$ and $\phi_n$ is the usual harmonic oscillator eigenstate.  From the energies, we see that for $\Delta>0$, the minimum of $n>1$ LL lies at $k_z=0$, while for $\Delta<0$, its minimum lies at the Weyl points.  Compared with the 3D conical WSMs, where the zeroth LL is linear and extends from the valence band to the conduction band~\cite{P.E.C.Ashby, Y.X.Wanga}, here the zeroth LL is parabolic and thus is quite different.  Depending on $\Delta$, the zeroth LL intersects ($\Delta<0$) or does not intersect ($\Delta>0$) the zero energy.  The characteristics of the LLs are clearly seen in Fig.~\ref{Fig4}.

In the following, we study the optical conductivities in order to find  signatures that can characterize different phases.  In this work, we focus on the real (absorption) part of the conductivity and consider zero temperature, $T=0$.

\section{interband optical conductivity}

In this section, we investigate the interband optical conductivity in the system without a magnetic field.  The optical conductivity is calculated from the linear-response Kubo's formula, 
\begin{align}
\sigma_\alpha(\omega)=&\frac{-i}{V}\sum_{s,s'}\sum_{\boldsymbol k}
\frac{f(\varepsilon_s)-f(\varepsilon_{s'})}
{\varepsilon_s-\varepsilon_{s'}}
\frac{|\langle\psi_s|J_\alpha|\psi_{s'}\rangle|^2} {\omega+\varepsilon_s-\varepsilon_{s'}+i\Gamma}, 
\label{Kubo1}
\end{align}
where $\alpha=x,z$ is the direction that the optical field acts on, $\omega$ is the photon frequency,  $V$ is the volume of the system, $\Gamma$ is the scattering rate, $f(x)$ is the Fermi-Dirac distribution function, and $J_\alpha=-ie[r_\alpha,H]$ is the current density operator.   As the configuration of the uniform electromagnetic fields are considered, the initial and final states have the same $\boldsymbol k$. 

\begin{figure}
	\includegraphics[width=8.2cm]{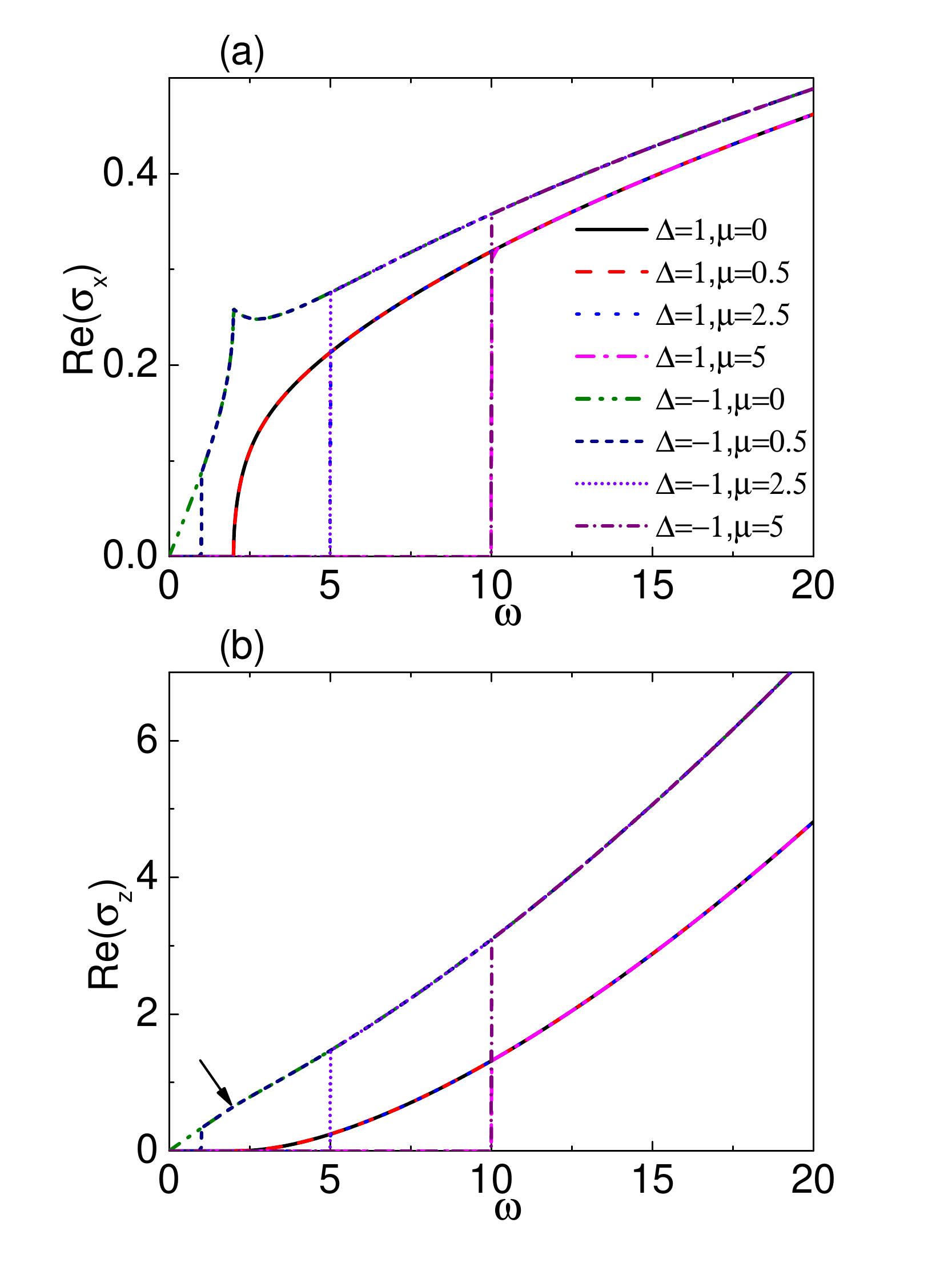}
	\caption{(Color online) Plot of the interband optical conductivity vs the photon frequency $\omega$ for different Fermi energy $\mu$ and $\Delta$ in the clean limit $\Gamma=0$.  (a) is for Re$(\sigma_x)$ (in unit of $\frac{\sigma_0}{\sqrt\zeta}$) and (b) is for Re$(\sigma_z)$ (in unit of $\frac{\sigma_0\sqrt\zeta}{v^2}$).  $\omega$ and $\mu$ are measured in units of $|\Delta|$.  In (b), the kink is labeled with arrow.  The legends are the same in both figures. }
	\label{Fig2}
\end{figure} 

In the clean limit $\Gamma=0$, the optical conductivity is determined by the photon absorption process, therefore the energy conversation must be satisfied, $\omega+\varepsilon_s=\varepsilon_{s'}$, requiring that the band indices are $s=-1$ and $s'=1$.  Thus $\varepsilon_{-1}=-\frac{\omega}{2}$, $\varepsilon_1=\frac{\omega}{2}$ and　$f(\varepsilon_{-1})-f(\varepsilon_1)=\theta(\frac{\omega}{2}+\mu)-\theta(-\frac{\omega}{2}+\mu)=\theta(\omega-2\mu)$, where $\mu$ is the Fermi energy and $\theta(x)$ denotes the step function.  After straightforward calculations, the analytical results for the conductivities at $\Gamma=0$ can be obtained~\cite{Supp}.  For $\Delta>0$, we have 

\begin{align}
\text{Re}(\sigma_x)=\frac{\sigma_03}{20\sqrt{2\zeta}}c_1
(1+\frac{4\Delta}{9\omega}+\frac{16\Delta^2}{9\omega^2})\theta(\omega-2\mu),
\label{sigmax1}
\end{align}
and
\begin{align}
\text{Re}(\sigma_z)=\frac{\sigma_0\sqrt{2\zeta}}{21 v^2}
c_1^3(1-\frac{6\Delta}{5\omega}-\frac{8\Delta^2}{5\omega^2})\theta(\omega-2\mu). 
\label{sigmaz1}   
\end{align}
For $\Delta<0$, we have 
\begin{align}
\text{Re}(\sigma_x)&=\frac{\sigma_03}{20\sqrt{2\zeta}}
\Big[
c_1(1+\frac{4\Delta}{9\omega}+\frac{16\Delta^2}{9\omega^2})
+c_2(-1+\frac{4\Delta}{9\omega}
\nonumber\\
&
-\frac{16\Delta^2}{9\omega^2})\theta(-\omega-2\Delta)
\Big]\theta(\omega-2\mu), 
\label{sigmax2}
\end{align}
and
\begin{align}
\text{Re}(\sigma_z)&=\frac{\sigma_0\sqrt{2\zeta}}{21v^2}
\Big[  
c_1^3(1-\frac{6\Delta}{5\omega}-\frac{8\Delta^2}{5\omega^2})
+c_2^3(-1-\frac{6\Delta}{5\omega}
\nonumber\\
&+\frac{8\Delta^2}{5\omega})\theta(-\omega-2\Delta)
\Big]\theta(\omega-2\mu).
\label{sigmaz2}
\end{align}
Here $\sigma_0=\frac{e^2}{2\pi}$ is the unit of the quantum conductivity, $c_1=\sqrt{\omega-2\Delta}$ and $c_2=\sqrt{-\omega-2\Delta}$.  It is interesting to find that Re$(\sigma_{x/z})$ has the same expression for the two cases of $\Delta>0$ and $0<-2\Delta<\omega$.  If we use $\frac{\sigma_0}{\sqrt\zeta}$ and $\frac{\sigma_0\sqrt\zeta}{v^2}$ as the unit of Re$(\sigma_x)$ and Re$(\sigma_z)$, respectively, the conductivities depend on the three quantities, $\Delta$, $\omega$ and $\mu$. 

It is worth emphasizing that in the band insulator phase $\Delta>0$, the result of Re$(\sigma_x)$ in Eq.~(\ref{sigmax1}) is quite different when compared with the previous work~\cite{E.Martino}.  In their work~\cite{E.Martino}, to explain the experimental data of optical conductivity, the authors proposed the 2D conical model to describe 3D ZrTe$_5$.  However, they made a slippery argument that only the behavior in the vicinity of the Weyl point was interested, and therefore, in the calculation of the current density operator, only the limiting region of  $\boldsymbol{k}\rightarrow0$ was considered.  As a result, the second and third term of Re$(\sigma_x)$ in Eq.~(\ref{sigmax1}) are missing in their work~\cite{E.Martino}, while these terms are actually very important when the photon frequency is comparable to $\Delta$, $\omega\sim\Delta$.  Our result is accurate and more reasonable by taking into account the contributions of all $\boldsymbol k$ in the whole Brillouin zone.  

We plot the interband optical conductivities Re$(\sigma_x)$ and Re$(\sigma_z)$ in Figs.~\ref{Fig2}(a) and (b), respectively.  First we consider the zero Fermi energy, $\mu=0$.  For the band insulator, the optical transitions are allowed only when the photon frequency is larger than the energy gap, $\omega>2\Delta$, and occur at the wave vector $k_z=\pm k_1= \pm\sqrt{-\frac{\Delta}{\zeta}+\frac{1}{2\zeta}\sqrt{\omega^2-\rho^2
}}$, with the condition $0<\rho=2v\sqrt{k_x^2+k_y^2}<\sqrt{\omega^2-4\Delta^2}$.  So in Fig.~\ref{Fig2}, there is a low-frequency cutoff in both Re$(\sigma_x)$ and Re$(\sigma_z)$.  Note that the contributions to the conductivity from the positive and negative wave vectors are equal.  For the WSM, the transitions at $k_z=\pm k_1$ are also allowed, but the condition changes as $0<\rho<\omega$.  Besides $k_z=\pm k_1$, the additional transitions at 
$k_z=\pm k_2 =\pm\sqrt{-\frac{\Delta}{\zeta}-\frac{1}{2\zeta} \sqrt{\omega^2-\rho^2}}$ are allowed, with the condition $0<\rho<\omega$ for 
$\omega<-2\Delta$ and $\sqrt{\omega^2-4\Delta^2}<\rho<\omega$ for $\omega>-2\Delta$.  As the contributions from $k_z=\pm k_2$ are different in the two regimes $\omega>-2\Delta$ and $\omega<-2\Delta$, in Fig.~\ref{Fig2}(a), there is a sharp increase in Re$(\sigma_x)$ when $\omega$ decreases to below $2\Delta$, and in Fig.~\ref{Fig2}(b), there is a kink in Re$(\sigma_z)$ at $\omega=2|\Delta|$ (shown with the arrow), with the first derivative being not continuous.  Therefore we obtain the distinct signatures of the optical conductivity at low frequency. 

\begin{figure}
	\includegraphics[width=8.2cm]{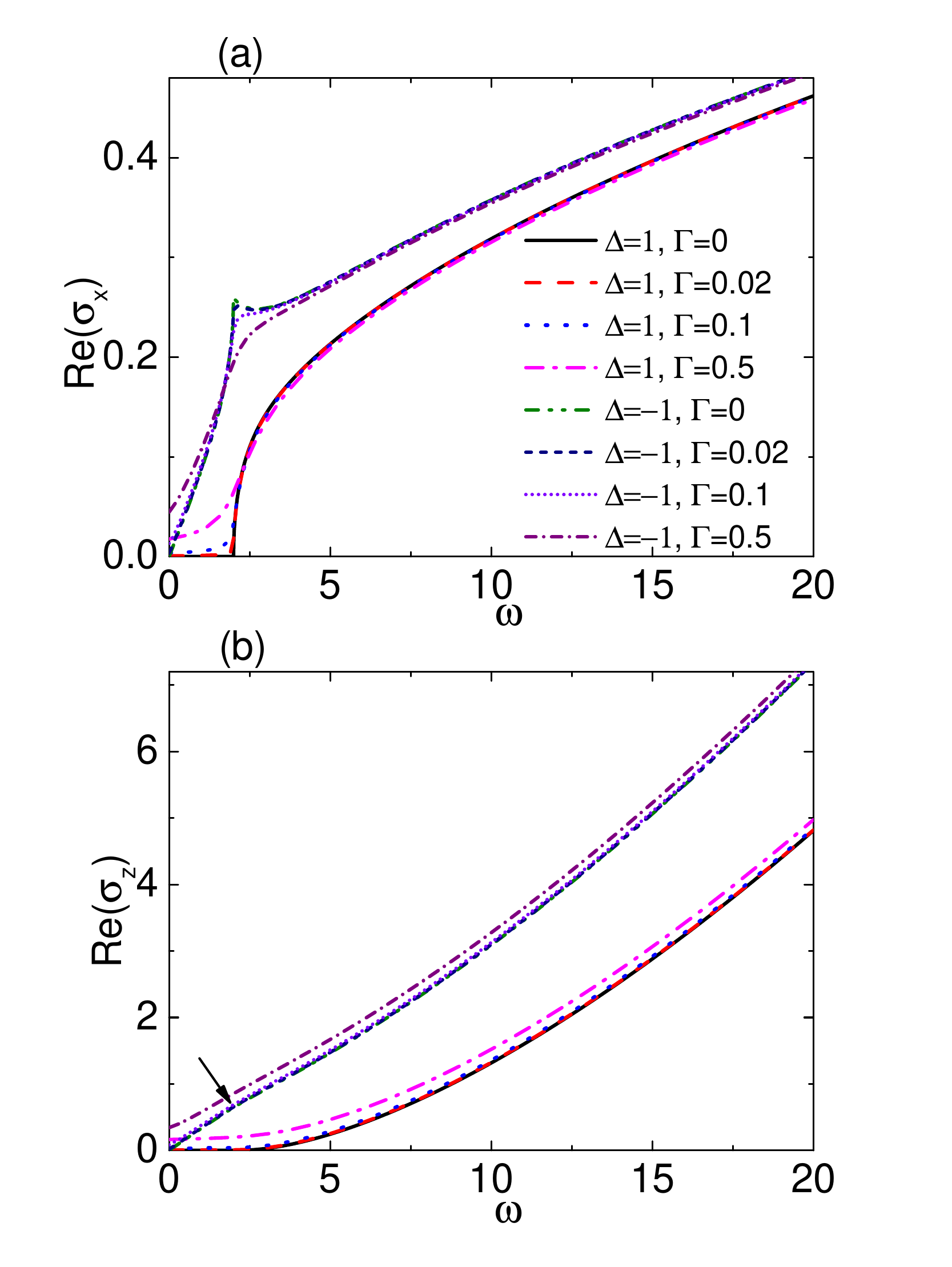}
	\caption{(Color online) Plot of the interband optical conductivity vs the photon frequency $\omega$ for different scattering rate $\Gamma$ and $\Delta$ with the Fermi energy $\mu=0$.  (a) is for Re$(\sigma_x)$ (in unit of $\frac{\sigma_0}{\sqrt\zeta}$) and (b) is for Re$(\sigma_z)$ (in unit of $\frac{\sigma_0\sqrt\zeta}{v^2}$).  $\omega$ and $\Gamma$ are measured in units of $|\Delta|$.  In (b), the kink is labeled with arrow.  The legends are the same in both figures. }
	\label{Fig3}
\end{figure} 

When the photon frequency is much larger than $|\Delta|$, $\omega\gg|\Delta|$, we can keep only the first term in Eqs.~(\ref{sigmax1})-(\ref{sigmaz2}) and neglect the other high-order terms.  For both $\Delta>0$ and $\Delta<0$, the asymptotic behaviors of the optical conductivities are obtained as,
\begin{align}
&\text{Re}(\sigma_x)\propto(\omega-2\Delta)^\frac{1}{2}\sim \omega^\frac{1}{2},
\label{asym1}
\\
&\text{Re}(\sigma_z)\propto(\omega-2\Delta)^\frac{3}{2}\sim\omega^\frac{3}{2},
\label{asym2}
\end{align}
The results show that for the 2D conical model, the asymptotic optical conductivity depends heavily on the direction that the optical field acts on, but is independent of the phase that the system lies in.  On one hand, the asymptotic Re($\sigma_x$) is consistent with the experimental results in ZrTe$_5$~\cite{E.Martino}, demonstrating the validity of the model in explaining the optical response.  On the other hand, to fully understand the topological nature of 3D ZrTe$_5$ and to evaluate the validity of the 2D conical model, further experimental studies are needed, especially of Re$(\sigma_z)$. 

For $d-$dimensional Dirac electrons, featuring the linear dispersions in all directions down to arbitrarily low energy, the interband optical response~\cite{P.Hosur2012, A.Bacsi} was found to scale as Re$(\sigma)\propto \omega^{(d-2)/z}$, with $z$ being the exponent in the band dispersion relation, $\varepsilon(k)\propto|k|^z$.  As the dispersion is isotropic, there is no difference for the optical response in different directions.  In 2D graphene, Re$(\sigma)$ is independent of the photon frequency, which has been verified in experiment~\cite{R.R.Nair, K.F.Mak}.  In 3D Dirac and Weyl semimetals, the optical conductivity is linear with the photon energy, which has been observed in Cd$_3$As$_2$~\cite{D.Neubauer, A.Akrap} and TaAs~\cite{B.Xu2016}.  For the 2D conical model studied in this work, the obtained exponents of $\frac{1}{2}$ and $\frac{3}{2}$ are quite different from the isotropic Dirac models.  

For the finite carrier density in real samples, the nonvanishing Fermi energy can screen the contributions to the conductivity from the conduction band that lies below it, $\varepsilon_1<\mu$.  As a result, the optical conductivity vanishes when $\omega<2\mu$ and the corresponding signatures are concealed, as shown in Fig.~\ref{Fig2}.  More importantly, when the Fermi energy satisfies $\mu>|\Delta|$, the optical conductivities exhibit similar shapes for $\Delta>0$ and $\Delta<0$. 　Therefore, it is unlikely to judge which phase the system lies in from the interband optical conductivity. 

The impurity scatterings are crucial in determining the transport properties of the system.  Here we include it phenomenologically by considering a nonvanishing scattering rate $\Gamma$ in Eq.~(\ref{Kubo1})~\cite{P.E.C.Ashby}.  The optical conductivity at finite $\Gamma$ is calculated numerically and the results are also plotted in Fig.~\ref{Fig3}.  We have carefully checked that as $\Gamma\rightarrow0$, the numerical results are consistent with the analytical ones.  We observe that for $\Delta>0$, the low-frequency cutoff in Re$(\sigma_x)$ and Re$(\sigma_z)$ will gradually disappear with increasing $\Gamma$.  At strong scattering rate $\Gamma=0.5$, the optical conductivity can reach a nonzero value even at zero frequency, $\omega=0$.  This is attributed to the fact that the increasing impurity scatterings can scatter more electronic states into the energy gap.  For $\Delta<0$, the sharp increase in Re$(\sigma_x)$ and the kink point in Re$(\sigma_z)$ are smoothened by the impurity scatterings.  However, the asymptotic behaviors of the optical conductivities are not affected by $\Gamma$, even when the impurity scatterings are sufficiently strong as $\Gamma=0.5\Delta$, which is accessible in the experimental condition.  For the mass parameter $\Delta=3$meV in ZrTe$_5$~\cite{E.Martino}, we estimate that the scattering rate $\Gamma=0.1\Delta=0.3$meV and the optical frequency $\omega=10\Delta=30$meV.

\section{Magneto-optical Conductivity}

In this section, we investigate the magneto-optical conductivity in the system when an external magnetic field is present, which can provide rich information about the LL structure and electron dynamics.  The previous experimental studies about magneto-optical spectroscopy in ZrTe$_5$ revealed that an exceptionally low magnetic field can drive the compound into the quantum limit~\cite{R.Y.Chenb, Z.G.Chen}.  The magneto-optical  conductivity can also be calculated from the Kubo's formula in Eq.~(\ref{Kubo1}).  As the quantized LLs are formed with the magnetic field, the Kubo's formula under the LL basis becomes
\begin{align}
\sigma_\alpha^B(\omega)&=\frac{-i}{2\pi l_B^2}\int_{-\infty}^\infty \frac{dk_z}{2\pi}\sum_{n,n'}\sum_{s,s'}
\frac{f(\varepsilon_{ns})-f(\varepsilon_{n's'})}
{\varepsilon_{ns}-\varepsilon_{n's'}}
\nonumber\\
&\times
\frac{|\langle\psi_{ns}|J_\alpha |\psi_{n's'}\rangle|^2}
{\omega+\varepsilon_{ns}-\varepsilon_{n's'}+i\Gamma}, 
\label{Kubo2}
\end{align}
where the factor $\frac{1}{2\pi l_B^2}=\frac{B}{\phi_0}$ denotes the degeneracy of each Landau state in an unit area of the $x-y$ plane.  For convenience, we use $\varepsilon_u=\frac{\sqrt2 v}{l_B}$ as the unit of energy and label the rescaled quantities as $\bar\varepsilon_{ns}=\frac{\varepsilon_{ns}}{\varepsilon_u}$, $\bar\Delta=\frac{\Delta}{\varepsilon_u}$, $\bar v=\frac{v}{\varepsilon_u}$, $\bar\zeta=\frac{\zeta}{\varepsilon_u}$, $\bar\mu=\frac{\mu}{\varepsilon_u}$, $\bar\omega=\frac{\omega}{\varepsilon_u}$ and $\bar\Gamma=\frac{\Gamma}{\varepsilon_u}$.  Here we choose the zero Fermi energy, $\bar\mu=0$, and focus only on the interband optical transitions. 

\begin{figure}
	\includegraphics[width=8.8cm]{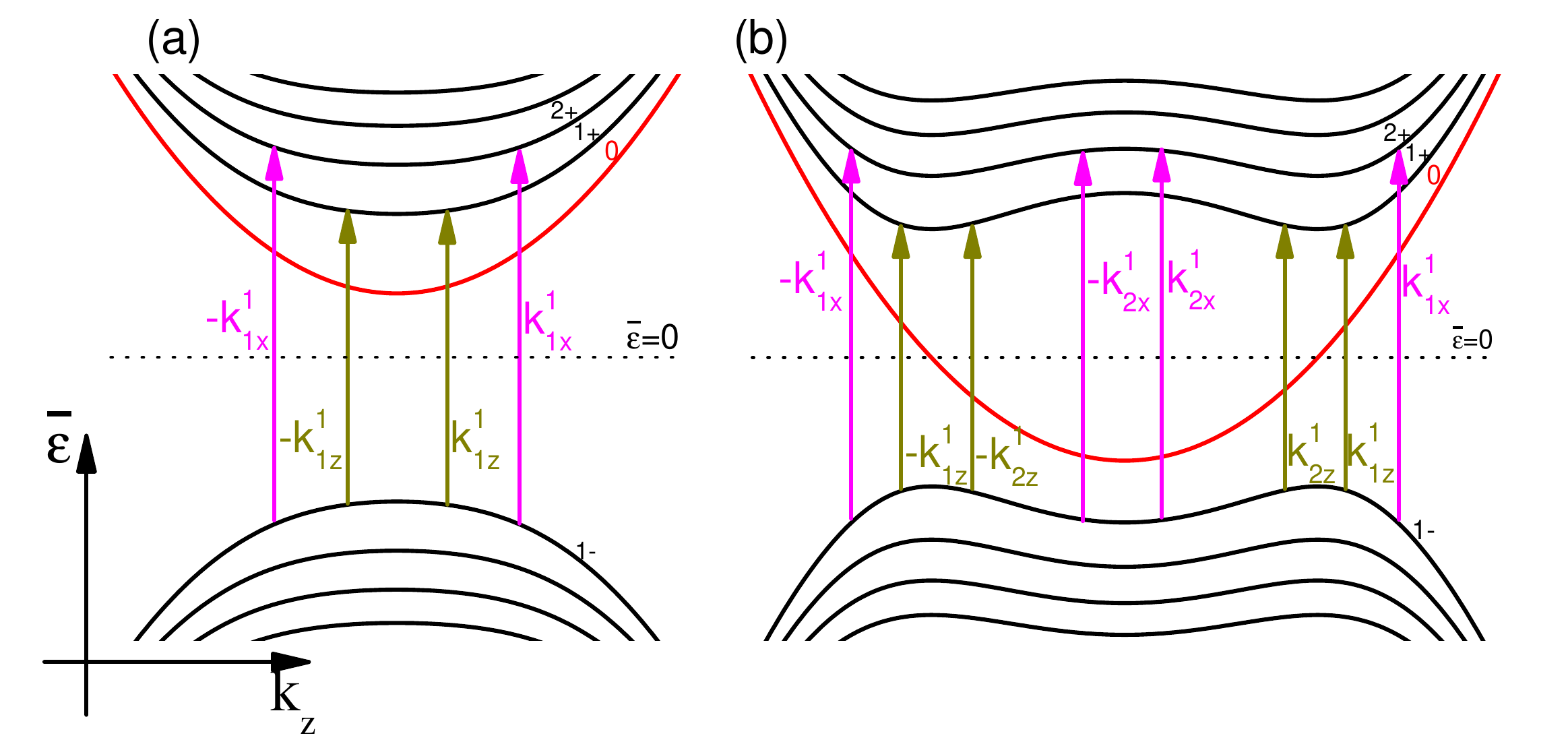}
	\caption{(Color online) Schematic plot of the dispersive LLs in (a) $\bar\Delta>0$ and (b) $\bar\Delta<0$.  The zeroth LL is shown in red color.  In both (a) and (b), the interband LL transitions of $1-\rightarrow 2+$ and $1-\rightarrow1+$ are indicated with arrows.  The wave vectors for the transitions are $k_{1x}^n=\sqrt{\frac{g_{n\lambda}}{2\bar\zeta}-\frac{\bar\Delta}{\bar\zeta}}$ and $k_{2x}^n=\sqrt{-\frac{g_{n\lambda}}{2\bar\zeta}-\frac{\bar\Delta}{\bar\zeta}}$, $k_{1z}^n=\sqrt{\frac{h_n}{2\bar\zeta}-\frac{\bar\Delta}{\bar\zeta}}$, $k_{2z}^n=\sqrt{-\frac{h_n}{2\bar\zeta}-\frac{\bar\Delta}{\bar\zeta}}$. }
	\label{Fig4}
\end{figure} 

In the clean limit $\bar\Gamma=0$, in addition to the energy conservation $\omega+\varepsilon_{ns}=\varepsilon_{n's'}$, the nonvanishing matrix element in Eq.~(\ref{Kubo2}) determines the optical selection rules between the initial and final states~\cite{P.E.C.Ashby, R.Y.Chenb, Y.X.Wanga, Y.X.Wangb, W.Duan}: if the optical field is perpendicular to the magnetic field, the Landau indices between the states differ by 1, $n-n'=\pm1$; if the optical field is parallel to the magnetic field, the Landau indices are the same, $n=n'$.  So the $n=0$ LL can contribute to Re$(\sigma_x^B)$, but not to Re$(\sigma_z^B)$.  The photon-assisted LL transitions are schematically plotted in Fig.~\ref{Fig4}, in which the wave vectors are explicitly indicated. 

In Re$(\sigma_z^B)$, where the parallel electric and magnetic fields are present, the intriguing phenomenon of the chiral anomaly arises~\cite{N.P.Armitage, P.Hosur2013}.  The chiral anomaly can lead to the nonconservative electron density at the two Weyl nodes and shift the local Fermi energy from zero to finite $\bar\mu_{+/-}$, with $\bar\mu_+=-\bar\mu_->0$~\cite{J.Behrends}.  However, if we assume that the variation of the local Fermi energy is below the $n=1$ LL,  $\bar\mu_+<\bar\varepsilon_{1+}$, the resonant peaks in Re$(\sigma_z^B)$ will not be affected by the chiral anomaly~\cite{Y.X.Wanga}.  

Using the obtained LL solutions, the magneto-optical conductivities can be derived at $\bar\Gamma=0$~\cite{Supp}.  For $\bar\Delta>0$, we have
\begin{align}
\text{Re}(\sigma_x^B)&=
\frac{\sigma_0}{8\sqrt{2\bar\zeta}\bar\omega^2}
\sum_{n\ge1}\sum_{\lambda=\pm1}
\frac{\bar\omega^2-\lambda\bar\omega g_n-2n-1}
{g_n\sqrt{g_n-2\bar\Delta}}
\nonumber\\
&+\frac{\sigma_0f_1}{4\sqrt{2\bar\zeta}\bar\omega}, 
\label{sigmax1B}
\end{align}
and 
\begin{align}
\text{Re}(\sigma_z^B)=\frac{\sigma_02\sqrt{2\bar\zeta}}
{\bar\omega^2l_B^2}\sum_{n\ge1}\frac{n\sqrt{h_n-2\bar\Delta}}{h_n}.   
\label{sigmaz1B}
\end{align}
For $\bar\Delta<0$, we have
\begin{align}
\text{Re}(\sigma_x^B)&=
\frac{\sigma_0}{8\sqrt{2\bar\zeta}\bar\omega^2}
\sum_{n\ge1}\sum_{\lambda=\pm1}
\Big(\frac{\bar\omega^2-\lambda\bar\omega g_n-2n-1} {g_n\sqrt{g_n-2\bar\Delta}}
\nonumber\\
&
+\frac{\bar\omega^2+\lambda\bar\omega g_n-2n-1}
{g_n\sqrt{-g_n-2\bar\Delta}}
\Big)
+\frac{\sigma_0}{4\sqrt{2\bar\zeta}\bar\omega}
\Big(f_1\theta(\bar\varepsilon_0>\bar\mu)
\nonumber\\
&
+f_2\theta(\bar\varepsilon_0<\bar\mu)
\Big), 
\label{sigmax2B}
\end{align}
and
\begin{align}
\text{Re}(\sigma_z^B)&=\frac{\sigma_02\sqrt{2\bar\zeta}}{\bar\omega^2l_B^2}
\sum_{n\ge1}\Big(\frac{n\sqrt{h_n-2\bar\Delta}}{h_n}
+\frac{n\sqrt{-h_n-2\bar\Delta}}{h_n}\Big). 
\label{sigmaz2B}
\end{align}
Here the parameters $f_1=\frac{1}{\sqrt{\bar\omega-\frac{1}{\bar\omega}-2\bar\Delta}}$,  $f_2=\frac{1}{\sqrt{\frac{1}{\bar\omega}-\bar\omega-2\bar\Delta}}$, $g_n=\sqrt{(\bar\omega-\frac{1}{\bar\omega})^2-4n}$ and $h_n=\sqrt{\bar\omega^2-4n}$.  In Eqs.~(\ref{sigmax1B}) and~(\ref{sigmax2B}) of Re$(\sigma_x^B)$, $\lambda=1$ and $-1$ represents $n-\rightarrow (n+1)+$ and $(n+1)-\rightarrow n-$ LL transition, respectively, and the last term(s) is related to $n=0$ LL transition.  In Eqs.~(\ref{sigmax2B}) and~(\ref{sigmaz2B}), the first term 
in the bracket corresponds to $k_z=\pm k_{1\alpha}^n$ transition and the second term corresponds to $k_z=\pm k_{2\alpha}^n$ transition.  If we use $\frac{\sigma_0}{\sqrt{\bar\zeta}}$ and $\frac{\sigma_0\sqrt{\bar\zeta}}{l_B^2}$ as the unit of Re$(\sigma_x^B)$ and Re$(\sigma_z^B)$, respectively, the magnetic-optical conductivities are also dependent on the three quantities, $\bar\Delta$, $\bar\omega$ and $\bar\mu$. 

\begin{figure}
	\includegraphics[width=8.8cm]{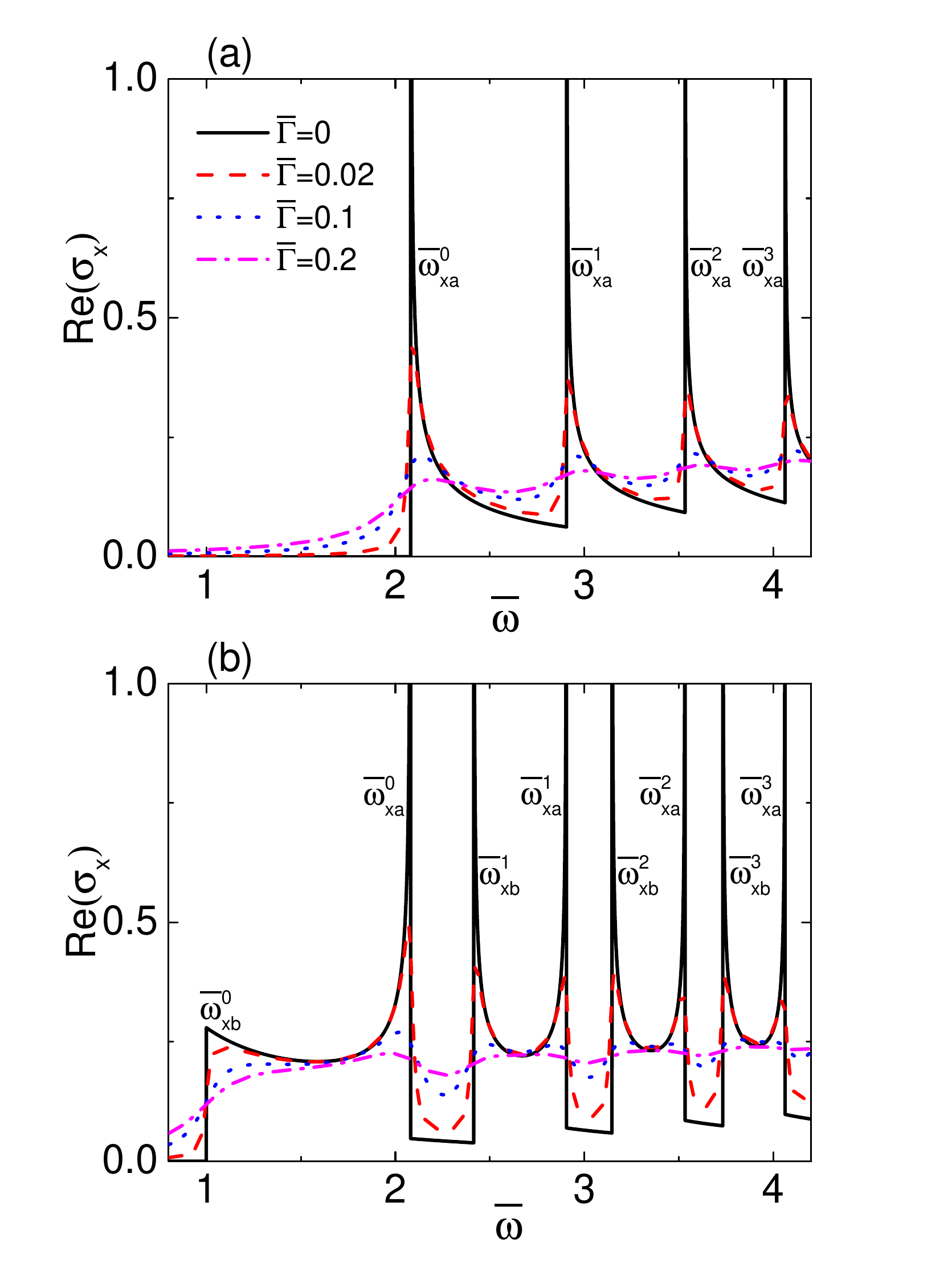}
	\caption{(Color online) The magneto-optical conductivity Re($\sigma_x^B$) (in unit of $\frac{\sigma_0}{\sqrt{\bar\zeta}}$) vs the photon frequency $\bar\omega$ for $\bar\Delta=0.8$ in (a) and $\bar\Delta=-0.8$ in (b).  $\omega$ and $\Gamma$ are measured in units of $\varepsilon_u$.  The characteristic frequencies $\bar\omega_{xa}^n$ and $\bar\omega_{xb}^n$ are indicated.  The results at finite $\bar\Gamma$ are also plotted and the legends are the same in both figures.  The Fermi energy is chosen as $\bar\mu=0$. }
	\label{Fig5}
\end{figure} 

\begin{figure}
	\includegraphics[width=8.8cm]{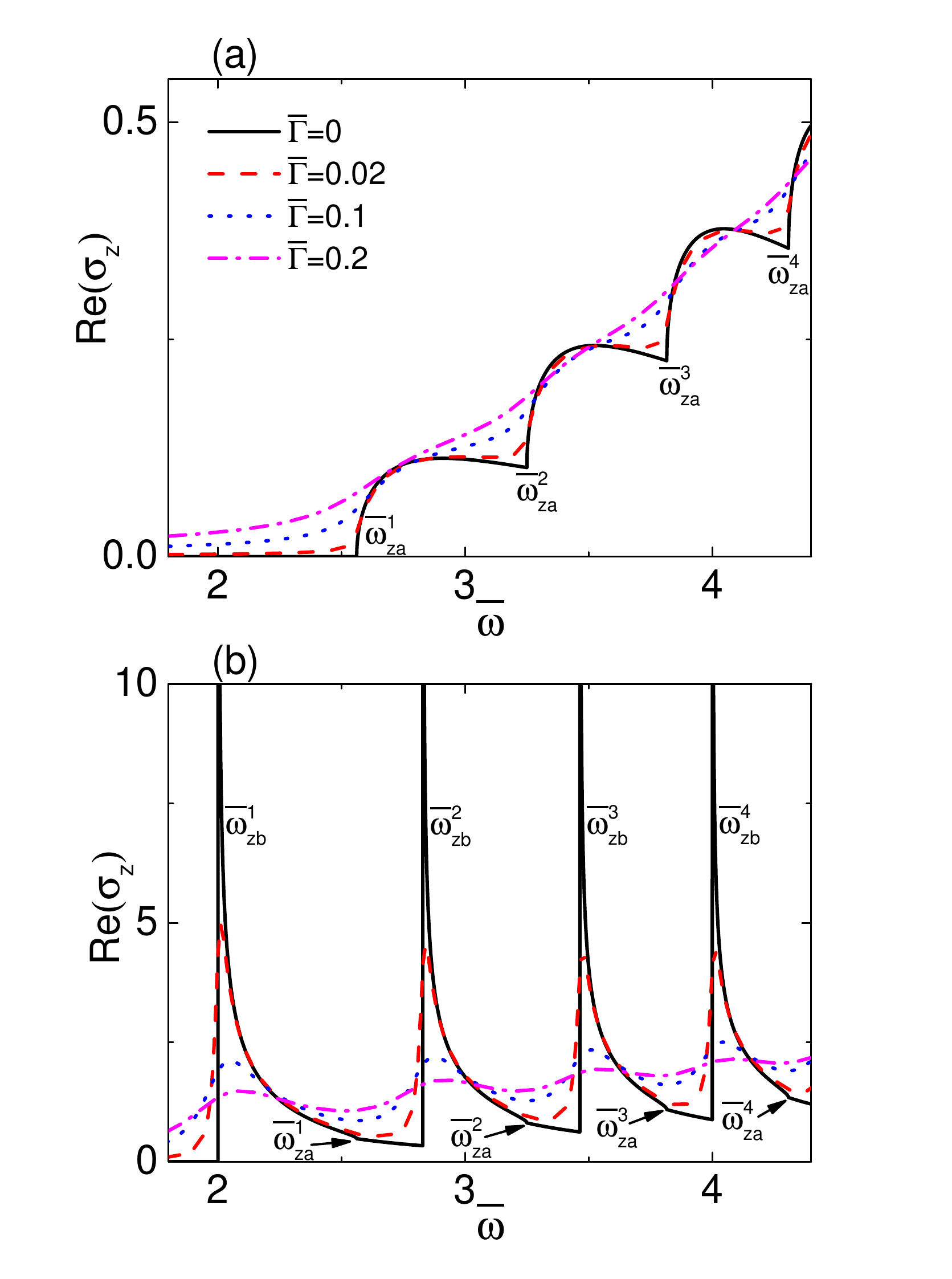}
	\caption{(Color online) The magneto-optical conductivity Re($\sigma_z^B$) (in unit of $\frac{\sigma_0\sqrt{\bar\zeta}}{l_B^2}$) vs the photon frequency $\bar\omega$ for $\bar\Delta=0.8$ in (a) and $\bar\Delta=-0.8$ in (b).  $\omega$ and $\Gamma$ are measured in units of $\varepsilon_u$.  The characteristic frequencies $\bar\omega_{za}^n$ and $\bar\omega_{zb}^n$ are indicated.  In (b), the kinks are shown with arrows.  The results at finite $\bar\Gamma$ are also plotted and the legends are the same in both figures.  The Fermi energy is chosen as $\bar\mu=0$.}
	\label{Fig6}
\end{figure}

The results of Re($\sigma_x^B$) and Re($\sigma_z^B$) are plotted as a function of $\bar\omega$ in Figs.~\ref{Fig5} and~\ref{Fig6}, respectively.  The resonant peaks in the conductivities correspond to the singularities in Eqs.~(\ref{sigmax1B})-(\ref{sigmaz2B}).  As the transitions persist with increasing frequency at larger $k_{1\alpha}^n$, the long tails of the peaks and the linear background in Re($\sigma_\alpha^B$) are thus induced.  This is consistent with the previous magneto-optical studies in 3D WSMs and other materials~\cite{P.E.C.Ashby, Y.X.Wanga, Y.X.Wangb}. 

First, we consider the conductivity in the band insulator, $\bar\Delta=0.8$.  For Re$(\sigma_x^B)$ in Fig.~\ref{Fig5}(a), with increasing frequency, the transition peaks appear.  One finds that both  $n-\rightarrow (n+1)+$ and $(n+1)-\rightarrow n+$ transitions have the onset frequency at $\bar\omega_{xa}^n=\sqrt{\bar\Delta^2+n}+\sqrt{\bar\Delta^2+n+1}$, which corresponds to a strong peak and is determined by $g_n=2\bar\Delta$ in Eq.~(\ref{sigmax1B}).  Note that $\bar\omega_{xa}^0=\bar\Delta+\sqrt{\bar\Delta^2+1}$ is a singularity of the last term.  For Re$(\sigma_z^B)$ in Fig.~\ref{Fig6}(a), there are weak peaks in the conductivity.  Intuitively, the energy difference between $n-$ and $n+$ LLs at $k_z=0$ gives the onset frequency at  $\bar\omega_{za}^n=2\sqrt{\bar\Delta^2+n}$.  But as $\bar\omega=\bar\omega_{za}^n$ is not a singularity of Eq.~(\ref{sigmaz1B}), it just gives the onset frequency of the transition and the weak peaks occur at a frequency larger than the onset frequency $\bar\omega>\bar\omega_{za}^n$. 

\begin{figure}
	\includegraphics[width=8.8cm]{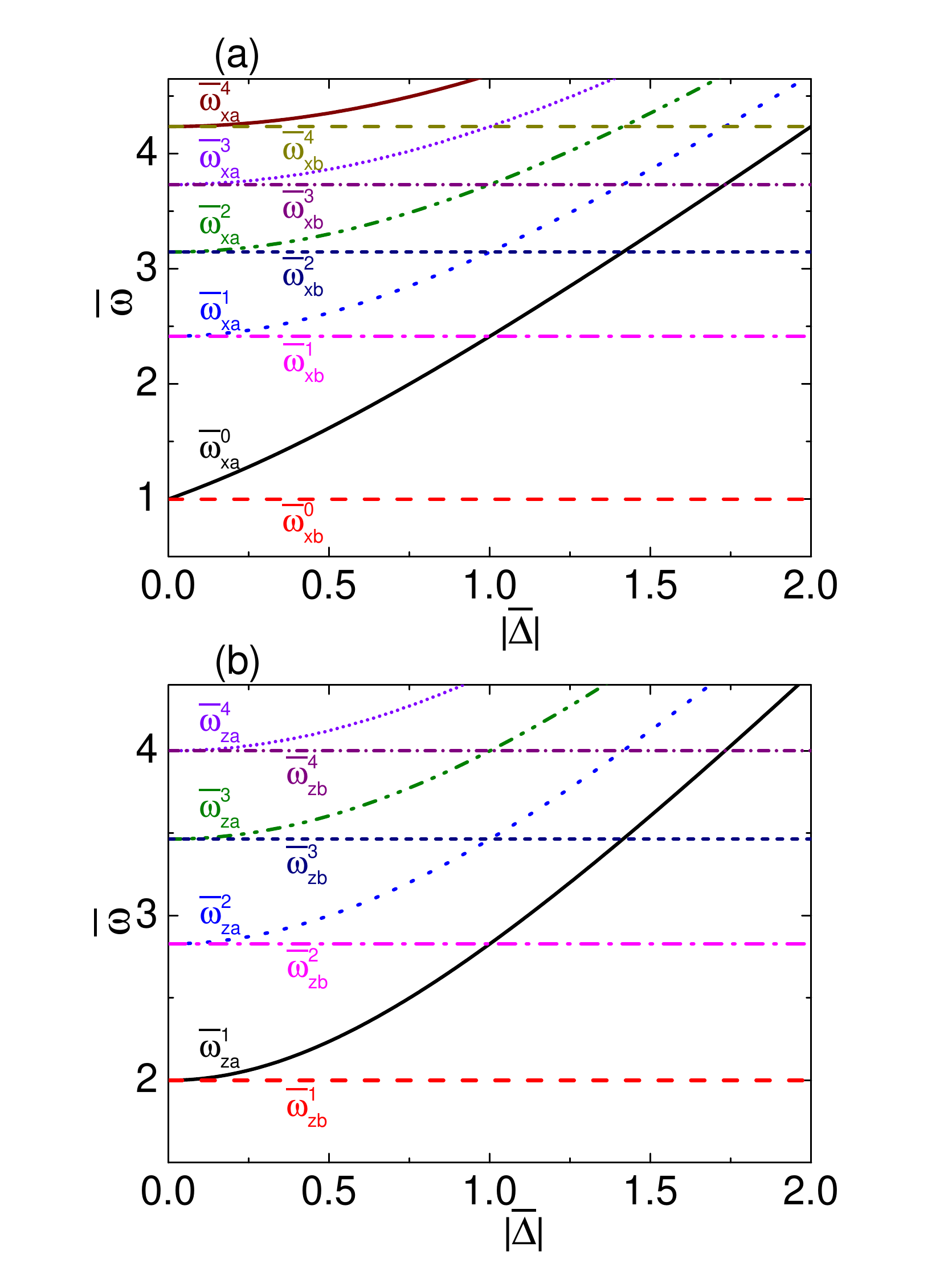}
	\caption{(Color online) Plot of the characteristic frequencies in the magneto-optical conductivity vs $|\bar\Delta|$.  $\omega$, $\omega_{\alpha a/b}^n$ and $\Delta$ are measured in units of  $\varepsilon_u$.  (a) $\bar\omega_{xa/b}^n$ for $0\le n\le4$ and (b) $\bar\omega_{za/b}^n$ for $1\le n\le4$. }
	\label{Fig7}
\end{figure} 

Next, we consider the WSM, $\bar\Delta=-0.8$.  For Re$(\sigma_x^B)$ in Fig.~\ref{Fig5}(b), two aspects are worth pointing out: (i) The double-strong peaks appear in the conductivity.  This is because four different photon absorptions are included in one LL transition, e.g., see $1-\rightarrow2+$ transition in Fig.~\ref{Fig4}(b).  Two are related to the wavevector $k_z=\pm k_{1x}^n$, with the onset frequency $\bar\omega_{xb}^n=\sqrt{n}+\sqrt{n+1}$, which is determined by $g_n=0$ in Eq.~(\ref{sigmax2B}).  Another two are related to $k_z=\pm k_{2x}^n$, and the corresponding transitions are restricted in a finite frequency range $\bar\omega_{xb}^n<\bar\omega<\bar\omega_{xa}^n$.  So there are no tails for these transitions.  The maximum frequency $\bar\omega_{xa}^n$ is determined by $g_n=2|\bar\Delta|$ in Eq.~(\ref{sigmax2B}) and is the same as the above $\bar\Delta>0$ case.  (ii) As the zeroth LL intersects the zero energy, it may lie in the conduction band or the valence band.  If $\bar\varepsilon_0>\bar\mu$, the $1-\rightarrow0$ transition occurs with $\bar\omega>\bar\omega_{xb}^0=1$, while if $\bar\varepsilon_0<0$, the $0\rightarrow1+$ transition is restricted in a finite frequency range $\bar\omega_{xb}^0<\bar\omega<\bar\omega_{xa}^0=-\bar\Delta+\sqrt{\bar\Delta^2+1}$.  Note that $\bar\omega=\bar\omega_{xa}^0$ is a singularity and corresponds to a strong peak, while $\bar\omega=\bar\omega_{xb}^0$ is not a singularity and there is no strong peak.  When the Fermi energy is tuned above (below) the zero energy, $\bar\mu>0$ ($\bar\mu<0$), the peak at $\bar\omega=\bar\omega_{xb}^0$ will move to high (low) frequency. 

For Re$(\sigma_z^B)$ in Fig.~\ref{Fig6}(b), there are strong peaks and kinks in the conductivity.  Similar to Re$(\sigma_x^B)$ in Fig.~\ref{Fig5}(b), four different photon absorptions are included in one transition, e.g., see $1-\rightarrow1+$ transition in Fig.~\ref{Fig4}(b).  Two are related to $k_z=\pm k_{1z}^n$ and occur at the onset frequency $\bar\omega_{zb}^n=2\sqrt n$, which is determined by $h_n=0$ of the first term in Eq.~(\ref{sigmaz2B}).  Another two are related to $k_z=\pm k_{2z}^n$ and are also restricted in a finite range $\bar\omega_{zb}^n<\bar\omega<\bar\omega_{za}^n$.  As the maximum frequency $\bar\omega=\bar\omega_{za}^n$ is not a singularity, it does not correspond to the strong peak in Re$(\sigma_z^B)$, but instead, to the kink, as shown by the arrows in Fig.~\ref{Fig6}(b).  

To explore the influence of $\Delta$ on the magneto-optical conductivity, we plot the characteristic frequencies $\bar\omega_{xa/b}^n$ and $\bar\omega_{za/b}^n$ as a function of $|\bar\Delta|$ in Fig.~\ref{Fig7}.  It is shown that $\bar\omega_{xa}^n$ and $\bar\omega_{za}^n$ increase with $|\bar\Delta|$, but $\bar\omega_{xb}^n$ and $\bar\omega_{zb}^n$ remain unchanged.  For the critical phase $\bar\Delta=0$, the characteristic frequencies overlap as $\bar\omega_{xa}^n=\bar\omega_{xb}^n$ and $\bar\omega_{za}^n=\bar\omega_{zb}^n$.  In the WSM phase, if $-1<\bar\Delta<0$, the double-strong peaks in Re$(\sigma_x^B)$ and the kinks in Re$(\sigma_z^B)$ are clearly resolved, which favor the observations in experiment.  If $\bar\Delta<-1$, we have $\bar\omega_{xa}^n>\bar\omega_{xb}^{n+1}$ and $\bar\omega_{za}^n>\bar\omega_{zb}^{n+1}$, meaning that the characteristic frequencies are mixed, which in turn makes them hard to be resolved in experiment. 

As the Fermi energy is finite in real samples, the low-frequency signatures of the magneto-optical conductivity will be screened by the Fermi surface, which is the same as the interband optical conductivity in the above section.  However, the high-frequency signatures still exist, as Re$(\sigma_x^B)$ exhibits strong resonant peaks in the gapped insulator phase and double-strong peaks in the WSM phase, while Re$(\sigma_z^B)$ exhibits weak peaks in the gapped insulator phase and the kink structure in the WSM phase.  Therefore we suggest that the magneto-optical conductivities can give clear evidences about which phase the system lies in.

We also consider the effect of the impurity scatterings on the magneto-optical conductivity phenomenologically, by including a finite scattering rate $\Gamma$ in Eq.~(\ref{Kubo2}).  The numerical results of the conductivity at finite $\bar\Gamma$ are plotted in Figs.~\ref{Fig5} and~\ref{Fig6}.  We have also carefully checked that as $\bar\Gamma\rightarrow0$, the numerical results are consistent with the analytical ones.  We see that for weak scattering $\bar\Gamma=0.02$, the resonant peaks in Re$(\sigma_x^B)$ and Re$(\sigma_z^B)$ are well preserved.  With increasing $\bar\Gamma$, the LL broadenings may be larger than the separations between neighboring LLs~\cite{J.Klier, Y.X.Wangc}.  Consequently, the impurity scatterings tend to blur out the resonant peaks in the conductivity~\cite{P.E.C.Ashby}.  In Figs.~\ref{Fig5} and~\ref{Fig6}, the resonant peaks disappear at strong scattering $\bar\Gamma=0.2$.  In Fig.~\ref{Fig6}(b), the kink structure in Re$(\sigma_z^B)$ at $\Delta<0$ is easily broken by the impurity scatterings, even when $\bar\Gamma$ is weak.  So the kink structure may not be easily discerned in experiment.  However, for weak $\bar\Gamma$, they can still provide important evidence to distinguish the two phases, since the weak peaks and strong peaks in the two different phase can be identified in Re$(\sigma_z^B)$.  Similar to disorder scattering, finite temperatures can also smooth out the absorption signatures~\cite{P.E.C.Ashby, Y.X.Wanga}.  Thus the key features in the magneto-optical conductivity would be observable as long as the temperature and the impurity scatterings are small compared to $\varepsilon_u$.  For the Fermi velocity $v=5\times10^5$m/s in ZrTe$_5$~\cite{E.Martino}, we estimate that $\varepsilon_u=18.1\sqrt B$meV.  So $\Delta=0.8\varepsilon_u$ used in Figs. 5(a) and 6(a) equals to the experimental value $\Delta=3$meV~\cite{E.Martino} when the magnetic field is weak as $B=0.043$T.  We also estimate that the photon frequency $\omega=2\varepsilon_u=36.2\sqrt B$meV, and a scattering rate $\Gamma=0.1\varepsilon_u= 1.81\sqrt B$meV.

\section{Conclusions} 

In conclusion, we have performed a systematic study on the frequency-dependent optical conductivity of the 2D conical model.  To evaluate the validity of this model in describing the topological properties of 3D ZrTe$_5$, we consider the optical conductivity in  both cases with and without an external magnetic field.  For the interband optical conductivities, the asymptotic behaviors at high frequency can support the validity of the 2D conical model.  The asymptotic behaviors also show robustness to the impurity scatterings.  For the magneto-optical conductivities, their distinct behaviors can give further information about which phase the system lies in.  As the impurity scatterings can smooth out the signatures in the magneto-optical conductivity, the clean ZrTe$_5$ samples are favored in experiment.  Our work helps characterize the low-energy electronic states in ZrTe$_5$ and further broaden the understanding of topological nature of the anisotropic Dirac electrons.

\section{Acknowledgments}

We would like to thank Biao Huang for many helpful discussions.  This work was supported by NSFC (Grants No. 11804122 and No. 11905054), and the Fundamental Research Funds for the Central Universities of China.

\end{document}


In this appendix, we give the detailed calculations for the optical conductivities of Eqs.~(8)-(11) and Eqs.~(15)-(18) in the main text.

\section{interband optical conductivity} 

The interband optical conductivity is calculated from the Kubo's formula 
\begin{align}
\sigma_\alpha
=-\frac{i}{V}\sum_{s,s'}\sum_{\boldsymbol k}
\frac{f(\varepsilon_s)-f(\varepsilon_{s'})}
{\varepsilon_s-\varepsilon_{s'}}
\frac{|\langle\psi_s|J_\alpha|\psi_{s'}\rangle|^2}
{\omega+\varepsilon_s-\varepsilon_{s'}+i0^+}, 
\end{align}
where $\alpha=x/z$ is the direction that the optical field acts on, $V$ is the volume of the system, $f(x)$ is the Fermi-Dirac distribution function, and $J_\alpha=-ie[r_\alpha,H]$ is the current density operator.  We take zero temperature and zero Fermi energy, and the indice $s=-1$ and $s'=1$.  We focus on the real (absorption) part of the conductivity and Eq.~(1) becomes
\begin{align}
\text{Re}(\sigma_\alpha)
=\frac{\pi}{\omega V}\sum_{\boldsymbol k}|\langle\psi_{-1}|J_{\boldsymbol k\alpha}|\psi_1\rangle|^2
\delta(\omega+\varepsilon_{-1}-\varepsilon_1). 
\end{align}
We assume that $2vk_x=x$, $2vk_y=y$, $\sqrt{2\zeta}k_z=z$, and change the Cartesian to the cylindrical system $x=\rho\text{cos}\theta$ and $y=\rho\text{sin}\theta$, then 
\begin{align}
\text{Re}(\sigma_\alpha)
&=\frac{\pi}{\omega V}
\sum_{\boldsymbol k}|\langle\psi_{-1}|J_{\boldsymbol k\alpha}|\psi_1\rangle|^2
\delta[\omega-2\sqrt{v^2(k_x^2+k_y^2)+(\Delta+\zeta k_z^2)^2}]
\nonumber\\
&=\frac{1}{32\sqrt{2\zeta}\pi^2 v^2\omega}
\iiint dxdydz
|\langle\psi_{-1}|J_\alpha|\psi_1\rangle|^2
\delta(\omega-\sqrt{x^2+y^2+(2\Delta+z^2)^2})
\nonumber\\
&=\frac{1}{16\sqrt{2\zeta}\pi v^2\omega}
\iiint \rho d\rho dz
|\langle\psi_{-1}|J_\alpha|\psi_1\rangle|^2
\delta(\omega-\sqrt{\rho^2+(2\Delta+z^2)^2}). 
\end{align}
To do the integration over $z$, we need to change the argument in the $\delta-$function by using the following formula,  
\begin{align}
\delta(\omega-\sqrt{\rho^2+(2\Delta+z^2)^2})
=\sum_i\frac{\delta(z-z_i)}{|d\sqrt{\rho^2+(2\Delta+z^2)^2}/dz|}, 
\end{align}
where the root $z_i$ is determined by the equation $\omega-\sqrt{\rho^2+(2\Delta+z^2)^2}=0$.  The roots are solved as  
\begin{align*}
&\text{if} \quad \Delta>0, \quad z_{1\pm}=\pm\sqrt{-2\Delta+\sqrt{\omega^2-\rho^2}}, 
\quad \text{for} \quad 0<\rho<\sqrt{\omega^2-4\Delta^2},
\quad \text{Case 1}
\\
&\text{if} \quad \Delta<0 \text{ and } -2\Delta>\omega, 
\quad \left\{ \begin{array}{l}
z_{1\pm}=\pm\sqrt{-2\Delta+\sqrt{\omega^2-\rho^2}},  
\quad \text{for} \quad 0<\rho<\omega,
\quad 
\text{Case 2}
\\ 
z_{2\pm}=\pm\sqrt{-2\Delta-\sqrt{\omega^2-\rho^2}},    
\quad \text{for} \quad 0<\rho<\omega, 
\quad \text{Case 3} 
\end{array} \right.
\\
&\text{if} \quad \Delta<0 \text{ and } -2\Delta<\omega, 
\quad \left\{ \begin{array}{l}
z_{1\pm}=\pm\sqrt{-2\Delta+\sqrt{\omega^2-\rho^2}},  
\quad \text{for} \quad 0<\rho<\omega,
\quad 
\text{Case 2}
\\ 
z_{2\pm}=\pm\sqrt{-2\Delta-\sqrt{\omega^2-\rho^2}},    
\quad \text{for} \quad \sqrt{\omega^2-4\Delta^2}<\rho<\omega, 
\quad \text{Case 4} 
\end{array} \right.
\end{align*}

\subsection{Calculation of $\sigma_x$}

For the conductivity along the $x-$direction, we have the matrix element of the current density operator $\langle\psi_{-1}|J_x|\psi_1\rangle=-ev(\frac{\Delta+\zeta k_z^2}{\varepsilon}\text{cos}\theta+i\text{sin}\theta)$.In Ref. [22], the authors considered the matrix element in the vicinity of the Weyl point, that is, in the limit $\boldsymbol k=0$, and obtained $|J_x|^2=e^2v^2$, as in (S38) in their Supplementary material. Then they used the simplified matrix element to calculate Re$(\sigma_x)$. Note that when $\boldsymbol k=0$, our expression of $|J_x|^2$ can be reduced to their's. Here, we calculate Re$(\sigma_x)$ by including the full k-dependent matrix element in a straightforward manner and we have
\begin{align}
\text{Re}(\sigma_x)=\frac{\sigma_0}{16\sqrt{2\zeta}\omega^3}
\iint \rho d\rho dz 
[(2\Delta+z^2)^2+\omega^2]
\delta[\omega-\sqrt{\rho^2+(2\Delta+z^2)^2}], 
\end{align}
with $\sigma_0=\frac{e^2}{2\pi}$ being the unit of the quantum conductivity. 

Case 1: 
\begin{align}
\text{Re}(\sigma_{x1})=&\frac{\sigma_0}{16\sqrt{2\zeta}\omega^3}
\int_0^{\sqrt{\omega^2-4\Delta^2}} \rho d\rho (2\omega^2-\rho^2)
\frac{\omega}{\sqrt{\sqrt{\omega^2-\rho^2}-2\Delta}
\sqrt{\omega^2-\rho^2}}
=\frac{\sigma_03}{20\sqrt{2\zeta}}\sqrt{\omega-2\Delta}
(1+\frac{4}{9}\frac{\Delta}{\omega}+\frac{16\Delta^2}{9\omega^2}). 
\end{align}
This is Eq.~(8) in the main text. 

Case 2: 
\begin{align}
\text{Re}(\sigma_{x2})=&\frac{\sigma_0}{16\sqrt{2\zeta}\omega^3}
\int_0^\omega\rho d\rho(2\omega^2-\rho^2) 
\frac{\omega}{\sqrt{\sqrt{\omega^2-\rho^2}-2\Delta}
\sqrt{\omega^2-\rho^2}}
\nonumber\\
=&\frac{\sigma_03}{20\sqrt{2\zeta}}
\Big[(\sqrt{\omega-2\Delta}-\frac{5}{6}\sqrt{-2\Delta})
+\frac{4\Delta}{9\omega}\sqrt{\omega-2\Delta}
+\frac{16\Delta^2}{9\omega^2}(\sqrt{\omega-2\Delta}-\sqrt{-2\Delta})
\Big]. 
\end{align}

Case 3:
\begin{align}
\text{Re}(\sigma_{x3})=&\frac{\sigma_0}{16\sqrt{2\zeta}\omega^3}
\int_0^{\omega} \rho d\rho (2\omega^2-\rho^2)
\frac{\omega}{\sqrt{-\sqrt{\omega^2-\rho^2}-2\Delta}
\sqrt{\omega^2-\rho^2}}
\nonumber\\
=&\frac{\sigma_03}{20\sqrt{2\zeta}}
\Big[(-\sqrt{-\omega-2\Delta}+\frac{5}{6}\sqrt{-2\Delta})
+\frac{4\Delta}{9\omega}\sqrt{-\omega-2\Delta}
+\frac{16\Delta^2}{9\omega^2}(-\sqrt{-\omega-2\Delta}+\sqrt{-2\Delta})
\Big]. 
\end{align}
 
Case 4:
\begin{align}
\text{Re}(\sigma_{x4})=&\frac{\sigma_0}{16\sqrt{2\zeta}\omega^3}
\int_{\sqrt{\omega^2-4\Delta^2}}^{\omega} 
\rho d\rho(2\omega^2-\rho^2)
\frac{\omega}{\sqrt{-\sqrt{\omega^2-\rho^2}-2\Delta}
\sqrt{\omega^2-\rho^2}}
=\frac{\sigma_0}{8\sqrt{\zeta}}\sqrt{-\Delta}
(1+\frac{32\Delta^2}{15\omega^2}). 
\end{align}

Thus we get the result for $\Delta<0$.  If $-2\Delta>\omega$, we have
\begin{align}
\text{Re}(\sigma_x)&=\text{Re}(\sigma_{x2})+\text{Re}(\sigma_{x3})
\nonumber\\
&=\frac{\sigma_03}{20\sqrt{2\zeta}}\Big[
(\sqrt{\omega-2\Delta}-\sqrt{-\omega-2\Delta})
+\frac{4\Delta}{9\omega}(\sqrt{\omega-2\Delta}+\sqrt{-\omega-2\Delta})
+\frac{16\Delta^2}{9\omega^2}(\sqrt{\omega-2\Delta}-\sqrt{-\omega-2\Delta})
\Big]. 
\end{align}
If $-2\Delta<\omega$, we have
\begin{align}
\text{Re}(\sigma_x)=\text{Re}(\sigma_{x2})+\text{Re}(\sigma_{x4})
=\frac{\sigma_03}{20\sqrt{2\zeta}}
\Big[\sqrt{\omega-2\Delta}
+\frac{4\Delta}{9\omega}\sqrt{\omega-2\Delta}
+\frac{16\Delta^2}{9\omega^2}\sqrt{\omega-2\Delta}
\Big]. 
\end{align}
The above results of Eqs.~(10) and (11) are included in Eq.~(10) in the main text.

\subsection{Calculation of $\sigma_z$}

For the conductivity along the $z-$direction, we have $\langle\psi_{-1}|J_z|\psi_1\rangle=\frac{2e\zeta k_zv}{\varepsilon}\sqrt{k_x^2+k_y^2}$, then 
\begin{align}
\text{Re}(\sigma_z)&=\frac{\sigma_0\sqrt\zeta}{4\sqrt2v^2\omega^3} 
\iint \rho d\rho dz \rho^2 z^2 
\delta[\omega-\sqrt{\rho^2+(2\Delta+z^2)^2}]. 
\end{align}

Case 1:
\begin{align}
\text{Re}(\sigma_{z1})&=\frac{\sigma_0\sqrt\zeta}{4\sqrt 2v^2\omega^3} 
\int_0^{\sqrt{\omega^2-4\Delta^2}} \rho d\rho  
\rho^2(\sqrt{\omega^2-\rho^2}-2\Delta)
\frac{\omega}{\sqrt{\sqrt{\omega^2-\rho^2}-2\Delta}\sqrt{\omega^2-\rho^2}}
\nonumber\\
&=\frac{\sigma_0\sqrt{2\zeta}}{21 v^2}(\omega-2\Delta)^\frac{3}{2}
(1-\frac{6\Delta}{5\omega}-\frac{8\Delta^2}{5\omega^2}). 
\end{align}
This is Eq.~(9) in the main text. 

Case 2:
\begin{align}
\text{Re}(\sigma_{z2})&=\frac{\sigma_0\sqrt\zeta}{4\sqrt 2v^2\omega^2} 
\int_0^{\omega} d\rho \rho^3  
\frac{\sqrt{\sqrt{\omega^2-\rho^2}-2\Delta}}{\sqrt{\omega^2-\rho^2}}
\nonumber\\
&=\frac{\sigma_0\sqrt{2\zeta}}{21v^2} 
\Big[
\omega\sqrt{\omega-2\Delta}
+\Delta(\frac{7}{2}\sqrt{-2\Delta}-\frac{16}{5}\sqrt{\omega-2\Delta})
+\frac{4\Delta^2}{5\omega}\sqrt{\omega-2\Delta}
-\frac{16\Delta^3}{5\omega^2}(\sqrt{-2\Delta}-\sqrt{\omega-2\Delta})
\Big]. 
\end{align}

Case 3:
\begin{align}
\text{Re}(\sigma_{z3})&=\frac{\sigma_0\sqrt\zeta}{4\sqrt 2v^2\omega}
\int_0^\omega \rho d\rho  
\frac{\rho^2(-\sqrt{\omega^2-\rho^2}-2\Delta)}{\omega^3}
\frac{\omega}{\sqrt{-\sqrt{\omega^2-\rho^2}-2\Delta}\sqrt{\omega^2-\rho^2}}
\nonumber\\
&=\frac{\sigma_0\sqrt{2\zeta}}{21v^2}
\Big[
\omega\sqrt{-\omega-2\Delta}
-\Delta(\frac{7}{2}\sqrt{-2\Delta}-\frac{16}{5}\sqrt{-\omega-2\Delta})
+\frac{4\Delta^2}{5\omega}\sqrt{-\omega-2\Delta}
+\frac{16\Delta^3}{5\omega^2}(\sqrt{-2\Delta}-\sqrt{-\omega-2\Delta})
\Big]. 
\end{align}
 
Case 4:  
\begin{align}
\text{Re}(\sigma_{z4})=&\frac{\sigma_0\sqrt\zeta}{4\sqrt 2v^2\omega^2}
\int_{\sqrt{\omega^2-4\Delta^2}}^{\omega} d\rho \rho^3  
\frac{\sqrt{-\sqrt{\omega^2-\rho^2}-2\Delta}}{\sqrt{\omega^2-\rho^2}}
=\frac{\sigma_0\sqrt\zeta}{105\sqrt 2v^2} 
\sqrt{-2\Delta}(\frac{32\Delta^3}{\omega^2}-35\Delta). 
\end{align}

Thus we get the result for $\Delta<0$.  If $-2\Delta>\omega$, we have
\begin{align}
\text{Re}{\sigma_z}=&\text{Re}(\sigma_{z2})+\text{Re}(\sigma_{z3})
\nonumber\\
=&\frac{\sigma_0\sqrt{2\zeta}}{21v^2}
\Big[
\omega(\sqrt{\omega-2\Delta}+\sqrt{-\omega-2\Delta})
-\frac{16\Delta}{5}(\sqrt{\omega-2\Delta}-\sqrt{-\omega-2\Delta})
+\frac{4\Delta^2}{5\omega}(\sqrt{\omega-2\Delta}+\sqrt{-\omega-2\Delta})
\nonumber\\
&+\frac{16\Delta^3}{5\omega^2}(\sqrt{\omega-2\Delta}-\sqrt{-\omega-2\Delta})
\Big]. 
\end{align}
If $-2\Delta<\omega$, we have
\begin{align}
\text{Re}(\sigma_z)=\text{Re}(\sigma_{z2})+\text{Re}(\sigma_{z4})
=&\frac{\sigma_0\sqrt{2\zeta}}{21v^2} 
\Big[
\omega\sqrt{\omega-2\Delta}
-\frac{16}{5}\Delta\sqrt{\omega-2\Delta}
+\frac{4\Delta^2}{5\omega}\sqrt{\omega-2\Delta}
+\frac{16\Delta^3}{5\omega^2}\sqrt{\omega-2\Delta}
\Big]. 
\end{align}
The above results of Eqs.~(17) and (18) are included in Eq.~(11) in the main text.

\section{magneto-optical conductivity}

The magneto-optical conductivity is also calculated from the Kubo's formula, which in the basis of the Landau Levels is given as:  
\begin{align}
\text{Re}(\sigma_\alpha^B)
&=\frac{1}{4\pi l_B^2\bar\omega}\sum_{n,n'}\sum_{s,s'}
\int_{-\infty}^\infty dk_z
|\langle\psi_{ns}|J_\alpha|\psi_{n's'}\rangle|^2
\delta(\bar\omega+\bar\varepsilon_{ns}-\bar\varepsilon_{n's'}). 
\end{align}
We take the indice as $s=-,s'=+$.

\subsection{Calculation of $\sigma_x^B$}

For the conductivity along the $x-$direction, we have the matrix element
$\langle n-|J_x|n'+\rangle
=\frac{el_B}{\sqrt2}(-\chi_{n+}\chi_{n'+}\delta_{n-1,n'}+\chi_{n-}\chi_{n'-}\delta_{n,n'-1})$,  
which determines the selection rule of $n\pm1=n'$. 

If $n+1=n'$, the conductivity is
\begin{align}
\text{Re}(\sigma_x^B)=\frac{\sigma_0}{4\bar\omega}
\sum_{n\ge1}\sum_i
\int_{-\infty}^\infty dk_z
(\frac{1}{2}-\frac{\bar\Delta+\bar\zeta k_z^2}{2\bar\varepsilon_n})
(\frac{1}{2}-\frac{\bar\Delta+\bar\zeta k_z^2}{2\bar\varepsilon_{n+1}})
\delta(k_z-k_i)
\frac{1}{|\frac{d(\bar\omega+\bar\varepsilon_{n-}-\bar\varepsilon_{n+1,+})}{dk_z}|}.
\end{align}
The roots of the equation $\bar\omega+\bar\varepsilon_{n-}-\bar\varepsilon_{n+1,+}=0$ are solved as, 
\begin{align*}
&\text{if}\quad \bar\Delta>0, \quad k_z=k_{z1\pm},
\quad \text{Case 1}
\\
&\text{if}\quad \bar\Delta<0, \quad
\left\{ \begin{array}{l}
k_z=k_{z1\pm},k_{z2\pm},   
\quad \text{for} \quad -\bar\Delta>\sqrt{(\frac{\bar\omega}{2}-\frac{1}{2\bar\omega})^2-n},
\quad 
\text{Case 2}
\\ 
k_z=k_{z1\pm}, \quad \text{for} \quad -\bar\Delta<\sqrt{(\frac{\bar\omega}{2}-\frac{1}{2\bar\omega})^2-n}, \quad \text{Case 3} 
\end{array} \right.
\\
&k_{z1\pm}=\pm\frac{1}{\sqrt{\bar\zeta}}
\sqrt{-\bar\Delta+\sqrt{(\frac{\bar\omega}{2}-\frac{1}{2\bar\omega})^2-n}}, 
\\
&k_{z2\pm}=\pm\frac{1}{\sqrt{\bar\zeta}}
\sqrt{-\bar\Delta-\sqrt{(\frac{\bar\omega}{2}-\frac{1}{2\bar\omega})^2-n}}. 
\end{align*}

For Cases 1 and 3, we have two roots $k_z=k_{z1\pm}$.  Then
\begin{align}
\text{Re}(\sigma_x^B)=\frac{\sigma_0}{8\sqrt{2\bar\zeta} \bar\omega^2}
\sum_{n\ge1}\frac{\bar\omega^2-\bar\omega \sqrt{(\bar\omega-\frac{1}{\bar\omega})^2-4n}-1-2n
}
{\sqrt{(\bar\omega-\frac{1}{\bar\omega})^2-4n}
\sqrt{\sqrt{(\bar\omega-\frac{1}{\bar\omega})^2-4n}-2\bar\Delta}
}. 
\end{align}

For Case 2, we have additional two roots $k_z=k_{k2\pm}$, which contribute to the conductivity as
\begin{align}
\text{Re}(\sigma_x^B)=\frac{\sigma_0}{8\sqrt{2\bar\zeta}\bar\omega^2}
\sum_{n\ge1}\frac{\bar\omega^2+\bar\omega \sqrt{(\bar\omega-\frac{1}{\bar\omega})^2-4n}-1-2n}
{\sqrt{(\bar\omega-\frac{1}{\bar\omega})^2-4n}
\sqrt{-\sqrt{(\bar\omega-\frac{1}{\bar\omega})^2-4n}-2\bar\Delta}
}. 
\end{align}

For the initial state $|1-\rangle$ and final state $|0\rangle$, we have the matrix element
$\langle 1-|J_x|0\rangle=-\frac{el_B}{\sqrt2}\chi_{1+}$
and 
\begin{align}
\text{Re}(\sigma_x^B)=\frac{\sigma_0}{4\bar\omega}
\sum_i\int_{-\infty}^\infty dk_z
(\frac{1}{2}+\frac{\bar\Delta+\bar\zeta k_z^2}{2\bar\varepsilon_1})
\delta(k_z-k_i)
\frac{1}{|\frac{d(\bar\omega+\bar\varepsilon_{1-}-\bar\varepsilon_0)}{dk_z}|}. 
\end{align}
The roots of $\bar\omega+\bar\varepsilon_{1-}-\bar\varepsilon_0=0$ are solved as  
$k_z=\pm\sqrt{\frac{1}{\bar\zeta}}\sqrt{\frac{\bar\omega}{2}-\frac{1}{2\bar\omega}-\bar\Delta}$.  Then 
\begin{align} 
\text{Re}(\sigma_x^B)
=\frac{\sigma_0}{4\sqrt{2\bar\zeta}\bar\omega}
\frac{1}{\sqrt{\bar\omega-\frac{1}{\bar\omega}-2\bar\Delta}}. 
\end{align}

If $n-1=n'$,
\begin{align}
\text{Re}(\sigma_x^B)=\frac{\sigma_0}{4\bar\omega}
\sum_{n\ge1}\sum_i\int_{-\infty}^\infty dk_z
(\frac{1}{2}+\frac{\bar\Delta+\bar\zeta k_z^2}{2\bar\varepsilon_{n+1}})
(\frac{1}{2}+\frac{\bar\Delta+\bar\zeta k_z^2}{2\bar\varepsilon_n})
\delta(k_z-k_i)
\frac{1}{|\frac{d(\bar\omega+\bar\varepsilon_{n-}-\bar\varepsilon_{n-1,+})}{dk_z}|}. 
\end{align}
The roots of the equation $\bar\omega+\bar\varepsilon_{n-}-\bar\varepsilon_{n-1,+}=0$ are solved as
\begin{align*}
&\text{if}\quad \bar\Delta>0, \quad k_z=k_{z1\pm},
\quad \text{Case 1}
\\
&\text{if}\quad \bar\Delta<0, \quad
\left\{ \begin{array}{l}
k_z=k_{z1\pm},k_{z2\pm},  
\quad \text{for} \quad -\bar\Delta>\sqrt{(\frac{\bar\omega}{2}-\frac{1}{2\bar\omega})^2-n}, 
\quad 
\text{Case 2}
\\ 
k_z=k_{z1\pm}, \quad \text{for} \quad -\bar\Delta<\sqrt{(\frac{\bar\omega}{2}-\frac{1}{2\bar\omega})^2-n}, \quad \text{Case 3} 
\end{array} \right.
\\
&k_{z1\pm}=\pm\frac{1}{\sqrt{\bar\zeta}}
\sqrt{-\bar\Delta+\sqrt{(\frac{\bar\omega}{2}-\frac{1}{2\bar\omega})^2-n}}, 
\\
&k_{z2\pm}=\pm\frac{1}{\sqrt{\bar\zeta}}
\sqrt{-\bar\Delta-\sqrt{(\frac{\bar\omega}{2}-\frac{1}{2\bar\omega})^2-n}}. 
\end{align*}

For Cases 1 and 3, we have two roots of $k_z=k_{z1\pm}$.  Then 
\begin{align}
\text{Re}(\sigma_x^B)=\frac{\sigma_0}{8\sqrt{2\bar\zeta}\bar\omega^2}
\sum_{n\ge1}
\frac{\bar\omega^2+\bar\omega\sqrt{(\bar\omega-\frac{1}{\bar\omega})^2-4n}-1-2n}
{\sqrt{(\bar\omega-\frac{1}{\bar\omega})^2-4n}
\sqrt{\sqrt{(\bar\omega-\frac{1}{\bar\omega})^2-4n}-2\bar\Delta}
}. 
\end{align}

For Case 2, we have the additional two roots of $k_z=k_{z2\pm}$, which contribute to the conductivity 
\begin{align}
\text{Re}(\sigma_x^B)=\frac{\sigma_0}{8\sqrt{2\bar\zeta}\bar\omega^2}
\sum_{n\ge1}\frac{\bar\omega^2-\bar\omega\sqrt{(\bar\omega-\frac{1}{\bar\omega})^2-4n}-1-2n}
{\sqrt{(\bar\omega-\frac{1}{\bar\omega})^2-4n}
\sqrt{-\sqrt{(\bar\omega-\frac{1}{\bar\omega})^2-4n}-2\bar\Delta}
}. 
\end{align}

For the initial state $|0\rangle$ and final state $|1+\rangle$, we have the matrix element
$\langle 0|J_x|1+\rangle=\frac{el_B}{\sqrt2}\chi_{1-}$ and 
\begin{align}
\text{Re}(\sigma_x^B)=\frac{\sigma_0}{4\bar\omega}
\sum_i\int_{-\infty}^\infty dk_z
(\frac{1}{2}-\frac{\bar\Delta+\bar\zeta k_z^2}{2\bar\varepsilon_1})
\delta(k_z-k_i)
\frac{1}{|\frac{d(\bar\omega+\bar\varepsilon_0-\bar\varepsilon_{1+})}{dk_z}|}. 
\end{align}
The roots of $\bar\omega+\bar\varepsilon_0=\bar\varepsilon_{1+}$ are solved as $
k_z=\pm\sqrt{\frac{1}{\bar\zeta}}\sqrt{\frac{1}{2\bar\omega}-\frac{\bar\omega}{2}-\bar\Delta}$.  Then 
\begin{align} 
\text{Re}(\sigma_x^B)=\frac{\sigma_0}{4\sqrt{2\bar\zeta}\bar\omega}
\frac{1}{\sqrt{\frac{1}{\bar\omega}-\bar\omega-2\bar\Delta}}. 
\end{align}
Eqs.~(21), (24) and (26) are included in Eq.~(15) in the main text. Eqs.~(21), (22), (24), (26), (27) and (29) are included in Eq.~(17) in the main text.

\subsection{Calculation of $\sigma_z^B$}

For the conductivity along the $z-$direction, we have the matrix element
$\langle n-|J_z|n'+\rangle
=2e\bar\zeta k_z(\chi_{n-}\chi_{n'+}\delta_{n,n'}
+\chi_{n+}\chi_{n'-}\delta_{n,n'})$,
which determines the selection rule $n=n'$.  Then the matrix element is
$\langle n-|J_z|n+\rangle=4e\bar\zeta k_z 
\chi_{n-}\chi_{n+}$, and
\begin{align}
\text{Re}(\sigma_z^B)=\frac{\sigma_0 8\bar\zeta^2}{\bar\omega l_B^2} 
\sum_{n\ge1} \sum_i
\int_{-\infty}^\infty dk_zk_z^2
\frac{n}{4\bar\varepsilon_n^2}
\delta(k_z-k_i)
\frac{1}{|\frac{d(\bar\omega+\bar\varepsilon_{n-}
-\bar\varepsilon_{n+})}{dk_z}|}. 
\end{align}
The roots of $\bar\omega+\bar\varepsilon_{n-}-\bar\varepsilon_{n+}=0$ as 
\begin{align*}
&\text{if}\quad \bar\Delta>0, \quad k_z=k_{z1\pm},
\quad \text{Case 1}
\\
&\text{if}\quad \bar\Delta<0, \quad
\left\{ \begin{array}{l}
k_z=k_{z1\pm},k_{z2\pm},  
\quad \text{for} \quad -\bar\Delta>\sqrt{\frac{\bar\omega^2}{4}-n}, 
\quad 
\text{Case 2}
\\ 
k_z=k_{z1\pm}, \quad \text{for} \quad -\bar\Delta<\sqrt{\frac{\bar\omega^2}{4}-n}, \quad \text{Case 3} 
\end{array} \right.
\\
&k_{z1\pm}=\pm\frac{1}{\sqrt{\bar\zeta}}
\sqrt{-\bar\Delta+\sqrt{\frac{\bar\omega^2}{4}-n}}
\\
&k_{z2\pm}=\pm\frac{1}{\sqrt{\bar\zeta}}
\sqrt{-\bar\Delta-\sqrt{\frac{\bar\omega^2}{4}-n}}
\end{align*}

For cases 1 and 3, we have two roots $k_z=k_{z1\pm}$.  Then
\begin{align}
\text{Re}(\sigma_z^B)
=\frac{\sigma_02\sqrt{2\bar\zeta}}{\bar\omega^2 l_B^2} 
\sum_{n\ge1}\frac{n\sqrt{\sqrt{\bar\omega^2-4n}-2\bar\Delta}}
{\sqrt{\bar\omega^2-4n}}. 
\end{align}
This is Eq.~(16) in the main text. 

For case 2, we have the additional roots $k_z=k_{z2\pm}$, which contribute to the conductivity
\begin{align}
\text{Re}(\sigma_z^B)
=\frac{\sigma_02\sqrt{2\bar\zeta}}{\bar\omega^2 l_B^2} 
\sum_{n\ge1}\frac{n\sqrt{-\sqrt{\bar\omega^2-4n}-2\bar\Delta}}
{\sqrt{\bar\omega^2-4n}}. 
\end{align}
Eqs.~(31) and (32) are included in Eq.~(18) in the main text.